\documentclass[twocolumn,a4paper,10pt]{article}
\usepackage{geometry}
\usepackage{amsmath,amssymb}
\usepackage{bm,braket}
\usepackage{multicol}
\usepackage{comment}
\usepackage[dvipdfmx]{graphicx}
\usepackage{cite}
\usepackage{authblk}
\usepackage{nidanfloat}

\title{{\Large Phase transition of an SU(3) symmetric spin-1 chain}}

\author{Tohru Mashiko and Kiyohide Nomura}
\affil{Department of Physics, Kyushu University, Fukuoka 819-0395, Japan}
\date{}

\begin{document}
\small
\twocolumn[
\maketitle
\begin{quotation}
We investigate a phase transition in an SU(3) symmetric spin-1 chain. 
To study this transition, we numerically diagonalize an SU(3) symmetric Hamiltonian combining the Uimin-Lai-Sutherland (ULS) model Hamiltonian with the Hamiltonian of the exact trimer ground state ($\mathbb{Z}_{3}$ symmetry breaking). 
Our numerical results are discussed on the basis of the conformal field theory (CFT) and the renormalization group (RG). 
We then show the phase transition between a trimer liquid phase (massless, no long-range order) and a trimer phase (spontaneously $\mathbb{Z}_{3}$ symmetry breaking).
Next, we find the central charge $c=2$ in the massless phase, and $c<2$ in the $\mathbb{Z}_{3}$ symmetry broken phase. 
From these results, we identify the universality class of the critical trimer liquid phase, as well as the phase boundary between these two phases.
\\
\end{quotation}
]

\section{Introduction}
\label{sec:intro}
In condensed matter physics, phase transitions and critical phenomena are important subjects.
In one- or two-dimensional quantum systems with continuous symmetry, quantum fluctuation is strong enough to prevent the long-range order, which makes critical phenomena complicated.
As for the SU(3) symmetric spin-1 chains, there have been several studies investigating the critical phenomena and the universality class. 
It was found that the ground state is the critical trimer liquid (TL) state with no long-range order, in the cases of the Uimin--Lai--Sutherland (ULS) model\cite{uimin,lai,sutherland,kulish} and the SU(3) symmetric Dimer--Trimer (DT) model\cite{oh,mashiko}.
In general, critical phases of SU($\nu$) symmetric quantum spin chains ($\nu$: integer) is stabilized by the $\mathbb{Z}_{\nu}$ symmetry\cite{affleck2,affleck3}, the center of the SU($\nu$) group.
In contrast, Hamiltonians describing the SU(3) trimer state were proposed\cite{schm,solyom,greiter}, where there is a trimer long-range order and the $\mathbb{Z}_{3}$ symmetry is spontaneously broken. 
However, there remain unsolved problems with regard to the phase transition between the TL phase and the trimer phase. 
Therefore, it is necessary to study such phenomena in more detail.
Revealing such phenomena gives basic theories to experiments and quantum simulations of ultracold alkaline earth metallic atoms in an optical lattice\cite{desalvo,gorshkov,taie}.

We here review an SU(3) symmetric spin-1 chain, the ULS model\cite{uimin,lai,sutherland,kulish}, to explain the critical phenomena of the TL. 
The ULS model is exactly solvable with the Bethe ansatz, and the ground state is the critical TL state. 
Affleck found\cite{affleck2,affleck3} that the universality class of the TL is the same as that of the level-1 SU(3) Wess--Zumino--Witten [SU$(3)_{1}$ WZW] model\cite{wess,witten,witten2} with the central charge $c=2$ and the scaling dimension $x=2/3$\cite{knizh}, by mapping the ULS model to the SU$(3)_{1}$ WZW model.
Also, he found\cite{affleck2,affleck3} the quasi-long-range order and soft modes at the wave number $q=0, \pm 2\pi/3$ in the TL state.
Then, Itoi and Kato calculated\cite{itoi} coefficients of logarithmic finite-size correction in the correlation functions in the case of the TL state.
They also found\cite{itoi} that the phase transition of the ULS model belongs to the Berezinskii--Kosterlitz--Thouless (BKT)-like universality class.

On the other hand, the pure trimer states, i.e., threefold degenerate states breaking translational invariance, were proposed\cite{nomura2,xian}. 
Here, the system follows periodic boundary conditions (PBCs). 
The pure trimer states on a spin-1 chain $\left| \psi_{\mathrm{T}} \right \rangle$ are given as
\begin{eqnarray}
	\left| \psi_{\mathrm{T}} \right \rangle = 
	\begin{cases}
		\{ \circ \circ \circ \} \{ \circ \circ \circ \} \cdots \{ \circ \circ \circ \} , \vspace{0.5em} \\
		\circ \{ \circ \circ \circ \} \cdots \{ \circ \circ \circ \} \circ \circ , \vspace{0.5em} \\
		\circ \circ \{ \circ \circ \circ \} \cdots \{ \circ \circ \circ \} \circ , \\
	\end{cases}
	\label{trimer}
\end{eqnarray}
where $\{ \circ \circ \circ \}$ is the singlet state of the three adjacent spins (trimer). 
The state of the trimer is written as
\begin{eqnarray}
	&&\{ \circ \circ \circ \} \equiv \frac{1}{\sqrt{6}} \left( |1,0,-1 \rangle + |0,-1,1 \rangle + |-1,1,0 \rangle \right. \notag \\
	&& \left. \hspace{5.5em} -|1,-1,0 \rangle - |0,1,-1 \rangle - |-1,0,1 \rangle \right), \label{trimer2}
\end{eqnarray}
where $1$, $0$, $-1$ refers to a spin magnetic quantum number $S^{z}$. 
Several models with the pure trimer ground states were proposed\cite{schm,solyom,greiter}, as a generalization of the Majumdar-Ghosh Hamiltonian\cite{majumdar} of the dimer ground state of the spin-1/2 chain. 
Also, there is a gap\cite{greiter} between threefold degenerate ground state energy and eightfold degenerate elementary excitation spectrum.

In this paper, we numerically diagonalize the Hamiltonian combining the ULS Hamiltonian\cite{uimin,lai,sutherland,kulish} $\hat{H}_{\mathrm{ULS}}$ with the trimer Hamiltonian\cite{greiter} $\hat{H}_{\mathrm{trimer}}$ under PBCs to investigate the phase transition between the TL phase and the trimer phase.
In Sec. \ref{sec:muls}, we define this Hamiltonian, which is composed only of exchange operators $\hat{P}_{ij}$. 
Numerical results of dispersion curves are shown in Sec. \ref{sec:dis}. 
In Sec. \ref{sec:tran}, we confirm that the phase transition is caused by a marginal operator\cite{itoi}, and then specify the universality class of the critical phenomena.
Conclusion and discussion are shown in Sec. \ref{sec:con}, where we discuss correlation functions describing the critical state of the TL and the long-range order of the trimer phase.

\section{MODEL}
\label{sec:muls}

To investigate the phase transition between the TL phase and the trimer phase, we numerically diagonalize the Hamiltonian defined as
\begin{eqnarray}
	\hat{H} \equiv \cos \phi \hat{H}_{\mathrm{ULS}} + \sin \phi \hat{H}_{\mathrm{trimer}}, \label{muls}
\end{eqnarray}
where the parameter $\phi$ is in the range of $0 \le \phi \le \pi/2$, and $\hat{H}_{\mathrm{ULS}}$ and $\hat{H}_{\mathrm{trimer}}$ will be defined in the next paragraph or later. 
In this section, we represent the Hamiltonian Eq. \eqref{muls} with the exchange operator $\hat{P}_{ii^{\prime}}$, which swaps spins at site $i$ with that at site $i^{\prime}$, defined as
\begin{eqnarray}
	\hat{P}_{ii^{\prime}} |\cdots S_{i}^{z} \cdots S_{i^{\prime}}^{z} \cdots \rangle = |\cdots S_{i^{\prime}}^{z} \cdots S_{i}^{z} \cdots \rangle, \label{p}
\end{eqnarray}
where $| \cdots \rangle$ is a state vector of a spin system and $S_{i}^{z}$ is a spin magnetic quantum number at site $i$.

First of all, we represent the ULS Hamiltonian with the exchange operator.
The Hamiltonian of the ULS model is defined\cite{uimin,lai,sutherland,kulish} as
\begin{eqnarray}
        \hat{H}_{\mathrm{ULS}} \equiv \sum_{i=1}^{N} \left[ \left(\hat{\bm{S}}_{i} \cdot \hat{\bm{S}}_{j} \right) + \left(\hat{\bm{S}}_{i} \cdot \hat{\bm{S}}_{j} \right)^{2} \right], \label{blbq}
\end{eqnarray}
where $\hat{\bm{S}}_{i}$ is a spin-1 operator at site $i$, and  we define $j \equiv i+1$.
Also, there is the relation\cite{sutherland} $\left(\hat{\bm{S}}_{i} \cdot \hat{\bm{S}}_{i^{\prime}} \right) + \left(\hat{\bm{S}}_{i} \cdot \hat{\bm{S}}_{i^{\prime}} \right)^{2} = \hat{P}_{ii^{\prime}} + \hat{1}$, where $\hat{1}$ is the identity operator.
Therefore, the ULS Hamiltonian can be rewritten\cite{sutherland} as
\begin{eqnarray}
        \hat{H}_{\mathrm{ULS}} = \sum_{i=1}^{N} \hat{P}_{ij}, \label{uls2}
\end{eqnarray}
where we neglect a trivial constant.

Secondly, we introduce the trimer Hamiltonian based on the paper of Greiter and Rachel\cite{greiter}.
Here, we denote $\hat{c}^{\dagger}_{i\sigma}$ ($\hat{c}_{i\sigma}$) to be an operator which creates (annihilates) a spin-1 with $S^{z}=\sigma$ at site $i$, and define the SU(3) generators at site $i$ as
\begin{eqnarray}
	&& \hat{J}^{a}_{i} \equiv \frac{1}{2} \sum_{\sigma,\sigma^{\prime}=1,0,-1} \hat{c}^{\dagger}_{i\sigma} \lambda^{a}_{\sigma\sigma^{\prime}} \hat{c}_{i\sigma^{\prime}}, \hspace{1em} a=1,\cdots,8 \label{gen} \\
	&& \hat{\bm{J}}_{i} \equiv \left(\hat{J}^{1}_{i}, \hat{J}^{2}_{i}, \hat{J}^{3}_{i}, \hat{J}^{4}_{i}, \hat{J}^{5}_{i}, \hat{J}^{6}_{i}, \hat{J}^{7}_{i}, \hat{J}^{8}_{i} \right)^{\mathrm{T}}, \label{genvec}
\end{eqnarray}
where the $\lambda^{a}$ are the Gell-Mann matrices (see Appendix \ref{subsec:gell}).
The operators  Eq. \eqref{gen} satisfy the commutation relations
\begin{eqnarray}
	\left[ \hat{J}^{a}_{i},\hat{J}^{b}_{i^{\prime}} \right] = \delta_{ii^{\prime}} f^{abc} \hat{J}^{c}_{i}, \hspace{2em} a,b,c=1,\cdots,8 
\end{eqnarray}
where $f^{abc}$ are the structure constants of SU(3) (see Appendix \ref{subsec:gell}), and we use the Einstein summation convention.
Then, we define $\hat{\bm{J}}^{(\nu)}_{i}$ with an integer $\nu$ for neighboring sites $i, \cdots, i+\nu-1$ as,
\begin{eqnarray}
	\hat{\bm{J}}^{(\nu)}_{i} \equiv \sum_{i^{\prime}=i}^{i+\nu-1} \hat{\bm{J}}_{i^{\prime}}.
\end{eqnarray}
For the trimer states Eq. \eqref{trimer2} on the four neighboring sites, the situation simplifies as we only have the two possibilities 
\begin{eqnarray}
	\begin{cases}
	\{ \circ \circ \circ\} \circ, \\
	\{ \circ \circ \} \{ \circ \circ \}.
	\end{cases}
	\label{auxi}
\end{eqnarray}
Here, $\{ \circ \circ \}$ is the triplet state of the two adjacent spins, whose state vectors are given by
\begin{eqnarray}
	\{ \circ \circ \} = 
	\begin{cases}
		\displaystyle{\frac{1}{\sqrt{2}}} (|1,0 \rangle - |0,1 \rangle), \vspace{0.5em} \\
		\displaystyle{\frac{1}{\sqrt{2}}} (|1,-1 \rangle - |-1,1 \rangle), \vspace{0.5em} \\
		\displaystyle{\frac{1}{\sqrt{2}}} (|0,-1 \rangle - |-1,0 \rangle).
	\end{cases}
	\label{triplet}
\end{eqnarray}
Note that Eqs. \eqref{auxi} and \eqref{triplet} are expressed\cite{greiter} by the representations of SU(3) (see Appendix \ref{subsec:rep}).
The quadratic Casimir operator for these two representations Eq. \eqref{auxi} has eigenvalues\cite{greiter} of 4/3 and 10/3 respectively.
Therefore, the auxiliary operators can be written as\cite{greiter}
\begin{eqnarray}
	\hat{H}_{i} = \left[\left(\hat{\bm{J}}^{(4)}_{i} \right)^{2} - \frac{4}{3} \right] \left[\left(\hat{\bm{J}}^{(4)}_{i} \right)^{2} - \frac{10}{3} \right]. \label{opetri}
\end{eqnarray}
The trimer Hamiltonian is defined as
\begin{eqnarray}
	\hat{H}_{\mathrm{trimer}} \equiv \sum_{i=1}^{N} \hat{H}_{i}. \label{hamitri}
\end{eqnarray}
Moreover, using the exchange operator $\hat{P}_{ii^{\prime}}$, there is an equation\cite{greiter} on the SU(3) generators as
\begin{eqnarray}
	\hat{\bm{J}}_{i} \hat{\bm{J}}_{i^{\prime}} = 
	\begin{cases}
	\displaystyle{\frac{4}{3}}, \hspace{7.2em} (i=i^{\prime}) \vspace{0.5em} \\
	\displaystyle{\frac{1}{2}} \left( \hat{P}_{ii^{\prime}} - \displaystyle{\frac{1}{3}} \right). \hspace{2em} (i \ne i^{\prime})
	\end{cases}
	\label{gen2}
\end{eqnarray}
Using Eqs. \eqref{opetri} -- \eqref{gen2}, the trimer Hamiltonian can be rewritten as
\begin{eqnarray}
	\hat{H}_{\mathrm{trimer}} &=& \sum_{i=1}^{N} \left[ \hat{P}_{ij} + \frac{2}{3}\hat{P}_{ik} + \frac{1}{3}\hat{P}_{il} + \hat{P}_{ij}\hat{P}_{ik} + \hat{P}_{ik}\hat{P}_{ij} \right. \notag \\
	&& \hspace{2.5em} + \frac{1}{2} \left( \hat{P}_{ij}\hat{P}_{il} + \hat{P}_{il}\hat{P}_{ij} +\hat{P}_{ik}\hat{P}_{il} + \hat{P}_{il}\hat{P}_{ik} \right) \notag \\
	&& \hspace{2.5em} \left. + \frac{1}{3} \left( \hat{P}_{ij}\hat{P}_{kl} + \hat{P}_{ik}\hat{P}_{jl} +\hat{P}_{il}\hat{P}_{jk} \right) \right], \label{trimerp}
\end{eqnarray}
where we define $k \equiv i+2$, $l \equiv i+3$.

From Eqs. \eqref{uls2} and \eqref{trimerp}, the Hamiltonian Eq. \eqref{muls} is composed only of the exchange operators, which conserves the number of spins, $N_{1}$, $N_{0}$, $N_{-1}$ for each state $S^{z}=1$, $0$, $-1$ respectively. 
Then, the full Hilbert space with $3^{N}$ dimensions is reduced to a subspace with $\frac{N!}{N_{1}!N_{0}!N_{-1}!}$ dimensions, $(N=N_{1}+N_{0}+N_{-1})$.

\section{Dispersion curves}
\label{sec:dis}
In this section, we show our numerical results of the dispersion curves of the model Eq. \eqref{muls}. 
Here, we make use of the conservation of the number of each spin mentioned in Sec. \ref{sec:muls} and the translational symmetry for our numerical calculations.  

Firstly, we let $\hat{O}_{t}$ denote a translational operator, which shifts all spins in the system by one site. $\hat{O}_{t}$ has an eigenvalue written as
\begin{eqnarray}
	\hat{O}_{t} |\cdots \rangle = e^{iq} |\cdots \rangle, \label{t}
\end{eqnarray}
where $q$ is the wave number. 
Under PBCs, $( \hat{O}_{t} )^{N}$ is an identity operator. 
The wave number thus should be $q=2\pi n/N$ ($n$: integer).

The energy eigenvalue $E$ is dependent on the wave number $q$ and the total spin quantum number of the system $S_{T}$. 
Therefore, we let $E_{S_{T}}(q)$ be the lowest energy at $q$ and $S_{T}$. 
We define the difference between $E_{S_{T}}(q)$ and the ground-state energy $E_{g}$ as
\begin{eqnarray}
	\Delta E_{S_{T}}(q) \equiv E_{S_{T}}(q) - E_{g}.
\end{eqnarray}
Also, we let $E(q)$ denote the lowest energy at $q$ and define the difference between $E(q)$ and $E_{g}$ as
\begin{eqnarray}
        \Delta E(q) \equiv E(q) - E_{g}. \label{eq}
\end{eqnarray}

\subsection{Dispersion curves in the case of the ULS model ($\phi=0$)}
\label{subsec:disuls}
\begin{figure}[t]
 \begin{center}
  \includegraphics[keepaspectratio,scale=0.23]{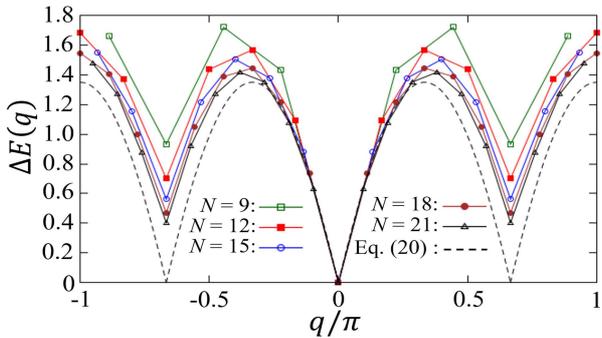}
 \end{center}
	\caption{Dispersion curves $\Delta E(q)$ in the case of $\phi=0$ with $N=9$--$21$. The dashed line is a curve obtained using Eq. \eqref{dis}.}
	\label{fig:eqph0}
\end{figure}

Figure \ref{fig:eqph0} shows dispersion curves $\Delta E(q)$ with $N=9$--$21$ in the case of $\phi=0$. 
The ground-state energy is the lowest energy at $q=0$ and $S_{T}=0$, that is, $E_{g} = E(0) = E_{0}(0)$. 
Also, there are soft modes at $q=0,\pm2\pi/3$ for all system sizes, as shown in Fig. \ref{fig:eqph0}. 
These results are consistent with the theory of Sutherland\cite{sutherland}. 
In this theory, $\Delta E(q)$ is given by
\begin{eqnarray}
	\Delta E(q) = D \left[\cos \left(\frac{\pi}{3} - |q| \right) - \frac{1}{2} \right], \hspace{0.5em} \left(0 \le |q| \le \frac{2 \pi}{3} \right) \notag \\ \label{dis} \\
	\Delta E(q) = \Delta E\left(|q| - \frac{2\pi}{3} \right), \hspace{0.5em} \left(\frac{2\pi}{3} \le |q| \le \pi \right) \notag 
\end{eqnarray}
in the thermodynamical limit, $N\rightarrow \infty$. 
Here, $D$ is a non-universal constant.
A dispersion curve gained from Eq. \eqref{dis} is also given in Fig. \ref{fig:eqph0}.
Furthermore, we reveal that $E(\pm 2\pi/3)=E_{1}(\pm 2\pi/3)=E_{2}(\pm 2\pi/3)$, namely, an eightfold degeneracy. 
Considering the fact that soft modes appear at $q=0, \pm 2\pi/3$, one should carry out numerical calculations only in cases where $N$ is a multiple of $3$ in later sections as well. 

\begin{figure}[t]
 \begin{center}
  \includegraphics[keepaspectratio,scale=0.23]{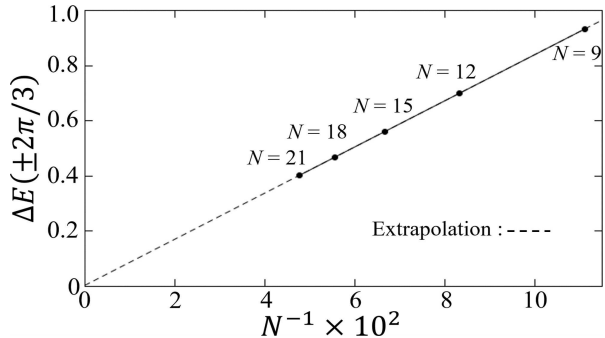}
 \end{center}
	\caption{Elementary excitation energy $\Delta E(\pm 2\pi/3)$ in the case of $\phi=0$ as a function of $N^{-1}$.}
        \label{fig:e-n}
\end{figure}

Figure \ref{fig:e-n} illustrates $\Delta E(\pm 2\pi/3)$ replotted for $N=9$--$21$.
The elementary excitation energy $\Delta E(\pm 2\pi/3)$ depends linearly on $N^{-1}$.
We extrapolate $\Delta E(\pm 2\pi/3)$ with the function $\Delta E(\pm 2\pi/3) = a_{0} + a_{1}N^{-1}$, where $a_{0}$ and $a_{1}$ are constants.
We then obtain $a_{0}=0.0045 \pm 0.0001$, and thus, it seems that the system is massless.

\subsection{Dispersion curves in the case of the trimer model ($\phi=\pi/2$)}
\label{subsec:distri}
\begin{figure}[t]
 \begin{center}
  \includegraphics[keepaspectratio,scale=0.2]{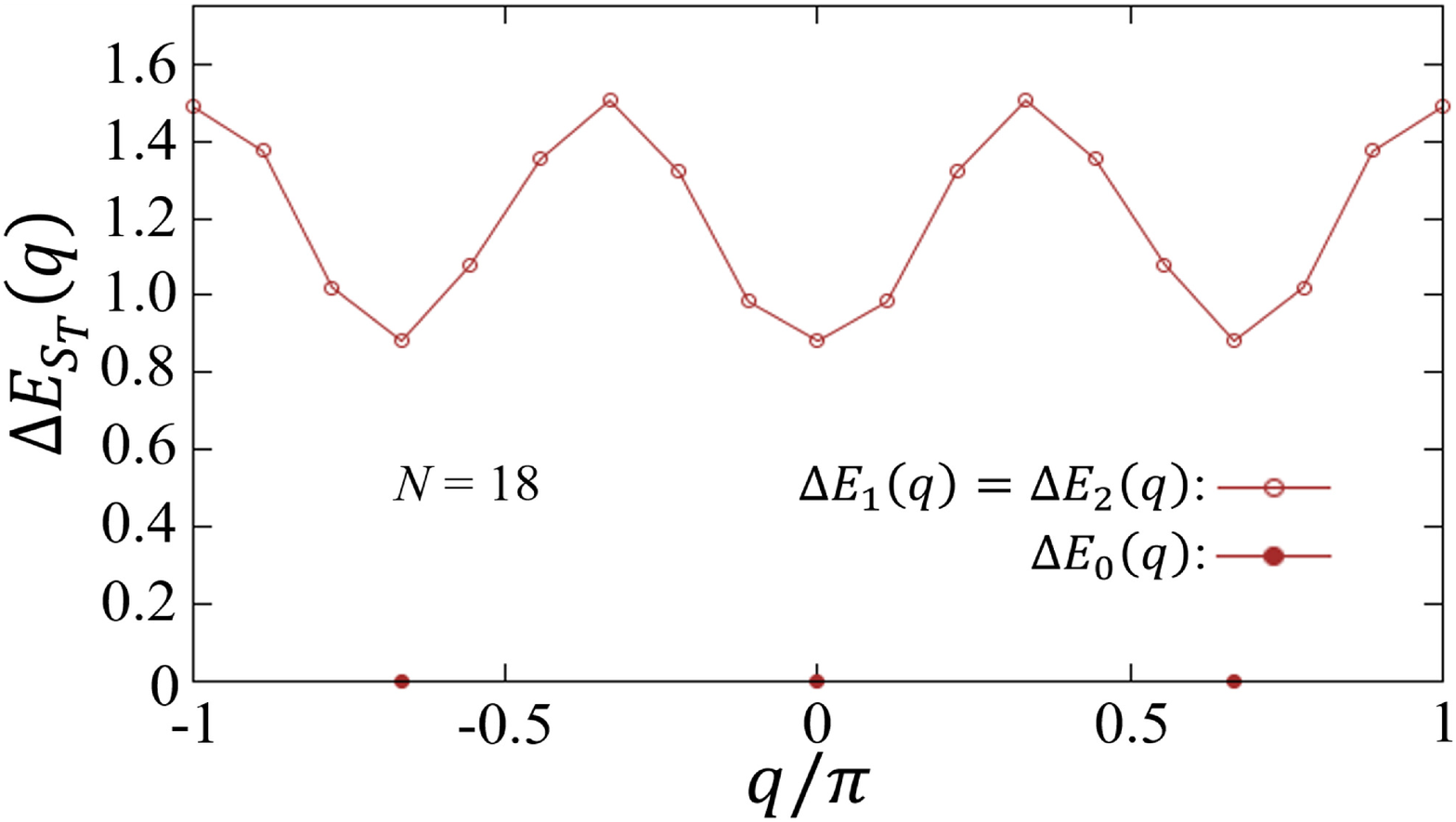}
 \end{center}
\begin{center}
  \includegraphics[keepaspectratio,scale=0.2]{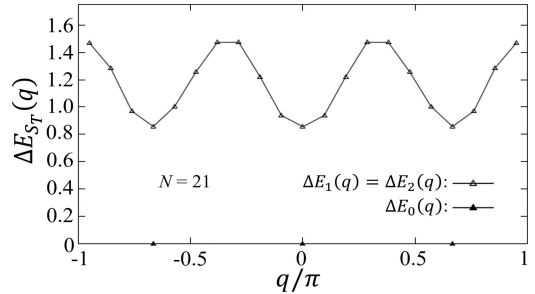}
 \end{center}
	\caption{Dispersion curves $\Delta E_{S_{T}}(q)$ in the case of $\phi=\pi/2$ with $N=15$--$21$.}
	\label{fig:eqph90}
\end{figure}

Figure \ref{fig:eqph90} shows dispersion curves $\Delta E_{S_{T}}(q)$ with $N=15$--$21$ in the case of $\phi=\pi/2$.
We find that the ground-state energy is threefold degenerated, $E_{g} = E_{0}(0) = E_{0}(\pm 2\pi/3)$.
We also find a gap between the $E_{g}$ and the eightfold degenerated elementary excitation spectrum, $\Delta E_{1}(q) =\Delta E_{2}(q)$.
According to Greiter and Rachel\cite{greiter}, elementary excitations are realized by the domain wall $\{\circ \circ\}$, that is 
\begin{eqnarray}
	\cdots \{ \circ \circ \circ\} \{\circ \circ\} \{ \circ \circ \circ\} \cdots .
\end{eqnarray}
If $N<15$, the domain wall appear only in the case of $N=3n-1$ ($n$: integer).
On the other hand, if $N \ge 15$, the domain wall can appear in the case of $N=3n$ as well, for example
\begin{eqnarray}
	\{ \circ \circ \circ\}\{\circ \circ\} \{ \circ \circ \circ\} \{\circ \circ\}\{ \circ \circ \circ\} \{\circ \circ\}, \hspace{0.8em} (N=15)
\end{eqnarray}
under PBCs.
Therefore, we believe that the excitations shown in Fig. \ref{fig:eqph90} is realized by the domain wall $\{\circ \circ\}$.
Moreover, the gap seems not to depend on the system size $N$, which is consistent with the paper of Greiter and Rachel\cite{greiter}.

\subsection{spin wave velocity}
\label{subsec:v}
\begin{figure}[t]
 \begin{center}
  \includegraphics[keepaspectratio,scale=0.23]{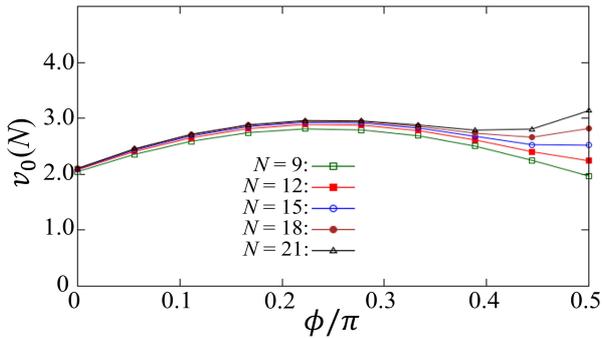}
 \end{center}
\caption{Spin wave velocity $v_{0}(N)$ as a function of $\phi$ with $N=9$--$21$.}
        \label{fig:v0}
\end{figure}

We calculate the spin wave velocity, which we utilize for later calculations of the central charge. 
Note that the spin wave velocity and the central charge are valid only in the case of the massless TL phase.
The spin wave velocity $v_{0}$ is defined as
\begin{eqnarray}
	v_{0} \equiv \left. \frac{dE(q)}{dq} \right|_{q=0}. \label{v0}
\end{eqnarray}
The velocity is a function of $N$, $v_{0}(N)$. 
In our numerical calculations, we investigate the slope of the dispersion curves to obtain $v_{0}(N)$ written as
\begin{eqnarray}
	v_{0}(N) = \frac{E(2\pi/N) - E(0)}{2\pi/N}. \label{v02}
\end{eqnarray}
The spin wave velocity is plotted in Fig. \ref{fig:v0}.
The area in the left side of Fig. \ref{fig:v0}, where size $N$ dependency of $v_{0}(N)$ is small, seems to be the TL phase, which we discuss in more detail in the next section.

\section{Phase transition and critical phenomena}
\label{sec:tran}

In this section, we investigate the phase transition between the TL phase and the trimer phase, which is caused by a marginal operator\cite{itoi}. 
We then calculate a critical exponent, the central charge $c$, to specify the universality class of the system.
Here, we define the difference between $E_{0}(q)$ and $E_{2}(q)$ as
\begin{eqnarray}
	\Delta E_{\mathrm{so}} \left( q \right) \equiv E_{0} \left( q \right) - E_{2} \left( q \right). \label{ese}
\end{eqnarray}

\subsection{Renormalization discussion on the SU$(3)_{1}$ WZW model}
\label{subsec:z3}

In this subsection, we investigate the phase transition characterized by $\mathbb{Z}_{3}$ symmetry breaking on the basis of the RG\cite{itoi}.

\begin{figure}[b]
 \begin{center}
  \includegraphics[keepaspectratio,scale=0.23]{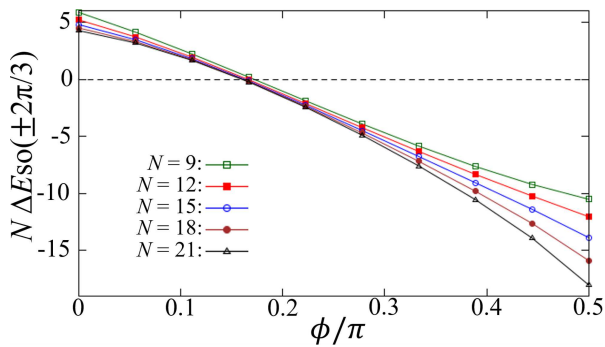}
 \end{center}
\caption{$N\Delta E_{\mathrm{so}}(\pm 2\pi/3)$ as a function of $\phi$ with $N=9$--$21$.}
        \label{fig:ese}
\end{figure}

\subsubsection{phase transition}
In Fig. \ref{fig:ese}, we plot $N\Delta E_{\mathrm{so}}(\pm 2\pi/3)$, for various $\phi$ with $N=9$--$21$.
As shown in Fig. \ref{fig:ese}, we firstly find that the value of $N\Delta E_{\mathrm{so}}(\pm 2\pi/3)$ changes from positive to negative at a certain point $\phi_{c}$ for all system sizes.
One can see that the size dependence of the crossing points $\phi_{c}(N)$ is very small, and that the slopes near $\phi_{c}(N)$ are almost the same.
On the contrary, there is relatively large size dependence far from $\phi_{c}(N)$.
We discuss these numerical results on the basis of the theory of Itoi and Kato\cite{itoi}. 
They analytically studied the action of the field in the vicinity of the system described by the SU$(3)_{1}$ WZW model, and then derived renormalization-group equations of the action. 
By solving the renormalization-group equations, they found\cite{itoi} that if the system is massless, $\Delta E_{\mathrm{so}}(\pm 2\pi/3)$ satisfies the relation
\begin{eqnarray}
	\Delta E_{\mathrm{so}} \left( \pm \frac{2\pi}{3} \right) > 0. \label{esep}
\end{eqnarray}
They also found\cite{itoi} that if the system is $\mathbb{Z}_{3}$ ordered, $\Delta E_{\mathrm{so}}(\pm 2\pi/3)$ satisfies the relation
\begin{eqnarray}
	\Delta E_{\mathrm{so}} \left( \pm \frac{2\pi}{3} \right) < 0. \label{esen}
\end{eqnarray}
By comparing our numerical results in Fig. \ref{fig:ese} with the theory\cite{itoi}, we find that the region $\phi<\phi_{c}$ is the $\mathbb{Z}_{3}$ symmetric (massless) phase, and the region $\phi>\phi_{c}$ is the $\mathbb{Z}_{3}$ ordered phase. 
In other words, there occurs a phase transition at $\phi=\phi_{c}$.

\subsubsection{phase transition and the renormalization group}
The phase transition shown in Fig. \ref{fig:ese} is described by the renormalization-group equation\cite{itoi} of the scaling variable $g_{1}$ for the marginal operator, Eq. \eqref{rg3}, (see Appendix \ref{sec:itoi}).
The solution of Eq. \eqref{rg3} is
\begin{eqnarray}
	g_{1}(l)= g_{1}(0) \left(1 - \frac{3}{2\sqrt{2}} g_{1}(0)l \right)^{-1}, \label{sol}
\end{eqnarray}
where $l \equiv \ln N$.
Here, $g_{1}(0)=0$ is a fixed point, which corresponds to the phase boundary $\phi_{c}$.
Also, in the vicinity of the fixed point, it is expected\cite{itoi,cardy1,cardy0.5} to be
\begin{eqnarray}
	g_{1}(0) = C (\phi - \phi_{c}), \label{g1ex}
\end{eqnarray}
where $C$ is a non-universal constant of proportionality.
We will discuss the expectation Eq. \eqref{g1ex} in more detail in Sec \ref{subsubsec:renvic}.
In the case of $g_{1}(0)<0$, $g_{1}(l)$ converges zero as $l \rightarrow \infty$ (marginally irrelevant), which means that the region $g_{1}(0)<0$ corresponds to the massless phase $\phi < \phi_{c}$.
In the case of $g_{1}(0)>0$, $g_{1}(l)$ diverges (marginally relevant), which means that the region $g_{1}(0)>0$ corresponds to the $\mathbb{Z}_{3}$ ordered phase $\phi > \phi_{c}$.
Now let us choose to halt the renormalization at the point where $g_{1}(l) = O(1)$ in the massive region.

Considering the marginal operator, the excitation energy $\Delta E_{S_{T}}(\pm 2\pi/3)$ is written\cite{itoi,cardy1,cardy0.5} as
\begin{eqnarray}
\Delta E_{S_{T}}\left(\pm \frac{2\pi}{3} \right) = \frac{2\pi v_{0}}{N} \left( \frac{2}{3} + d_{S_{T}} g_{1}(l) \right), \label{estp}
\end{eqnarray}
where $d_{S_{T}}$ is a coefficient depending on $S_{T}$ with the values\cite{affleck2,affleck3,itoi} of
\begin{eqnarray}
	d_{0} = -\frac{4}{3\sqrt{2}}, \hspace{2em} d_{1} = d_{2} = \frac{1}{6\sqrt{2}}. \label{cst}
\end{eqnarray}
Therefore, from Eqs. \eqref{ese} and \eqref{estp}, we obtain
\begin{eqnarray}
	N\Delta E_{\mathrm{so}} \left( \pm \frac{2\pi}{3} \right) = - \frac{3\pi}{\sqrt{2}} v_{0}(N) g_{1}(l). \label{eso3}
\end{eqnarray}

\subsubsection{massless phase}
As shown in Fig. \ref{fig:ese}, in the massless phase ($g_{1}(0)<0$) near $\phi=0$, $N\Delta E_{\mathrm{so}}(\pm 2\pi/3)$ gets smaller depending on $N$.
We discuss this fact on the basis of the renormalization group\cite{itoi,cardy1}.

In the massless phase near $\phi=0$, the limit of $g_{1}(0) \ll -1$, Eq. \eqref{sol} can be approximated as
\begin{eqnarray}
	g_{1}(l) \approx -\frac{2\sqrt{2}}{3} \cdot \frac{1}{\ln N}, \label{sol4}
\end{eqnarray}
in the SU(3) symmetric critical system.
From Eqs. \eqref{eso3} and \eqref{sol4}, we obtain
\begin{eqnarray}
       N\Delta E_{\mathrm{so}} \left( \pm \frac{2\pi}{3} \right) \approx \frac{2\pi v_{0}(N)}{\ln N}. \label{eso2}
\end{eqnarray}
Therefore, $N\Delta E_{\mathrm{so}} \left( \pm 2\pi/3 \right)$ in Fig. \ref{fig:ese} becomes smaller proportional to $1/\ln N$, since $v_{0}(N)$ is not very dependent on $N$ as shown in Fig. \ref{fig:v0}.

\subsubsection{$\mathbb{Z}_{3}$ ordered phase}
In the $\mathbb{Z}_{3}$ ordered phase ($g_{1}(0)>0$), $N\Delta E_{\mathrm{so}}(\pm 2\pi/3)$ negatively increases depending on $N$ in the vicinity of $\phi = \pi/2$, as shown in Fig. \ref{fig:ese}.
This fact can also be discussed on the basis of the renormalization group\cite{itoi,cardy1}.
If $g_{1}(0)>0$, $g_{1}(l)$ monotonically increases depending on $l$, as shown in Eq. \eqref{sol}.
Therefore, Eq.\eqref{eso3} shows that $N\Delta E_{\mathrm{so}}(\pm 2\pi/3)$ negatively increases as $N$ increases.
This means that in the thermodynamical limit, $E_{0}(\pm 2\pi/3)$ degenerate with the ground state energy $E_{g}$, which corresponds to the $\mathbb{Z}_{3}$ symmetry breaking (the trimer state).

Next, we discuss the correlation length in the $\mathbb{Z}_{3}$ ordered phase.
Here, we put $g_{1}(l^{\prime}) = 1$, where $l^{\prime}$ is the value of $l$ when the scaling variable is sufficiently renormalized.
Substituting $g_{1}(l^{\prime}) = 1$ for Eq. \eqref{sol}, we obtain
\begin{eqnarray}
	g_{1}(0) = \left( 1 + \frac{3}{2\sqrt{2}} l^{\prime} \right)^{-1}. \label{sol2}
\end{eqnarray}
Also, because of $\xi \sim e^{l^{\prime}}$, the correlation length can be written
\begin{eqnarray}
	\xi \sim \exp \left( \frac{2\sqrt{2}}{3} \frac{1 - g_{1}(0)}{g_{1}(0)} \right). \label{xi}
\end{eqnarray}
If $g_{1}(0) \rightarrow 0$, we obtain
\begin{eqnarray}
        \xi \sim \exp \left( \frac{2\sqrt{2}}{3} \frac{1}{g_{1}(0)} \right). \label{xi2}
\end{eqnarray}
In other words, the correlation length diverges as $\phi$ approaches the transition point, and the function form resembles that of the BKT-like\cite{itoi}.

\subsection{Renormalization discussion in the vicinity of $\phi=\phi_{c}$}
\label{subsec:z3vic}
\begin{figure}[b]
 \begin{center}
  \includegraphics[keepaspectratio,scale=0.2]{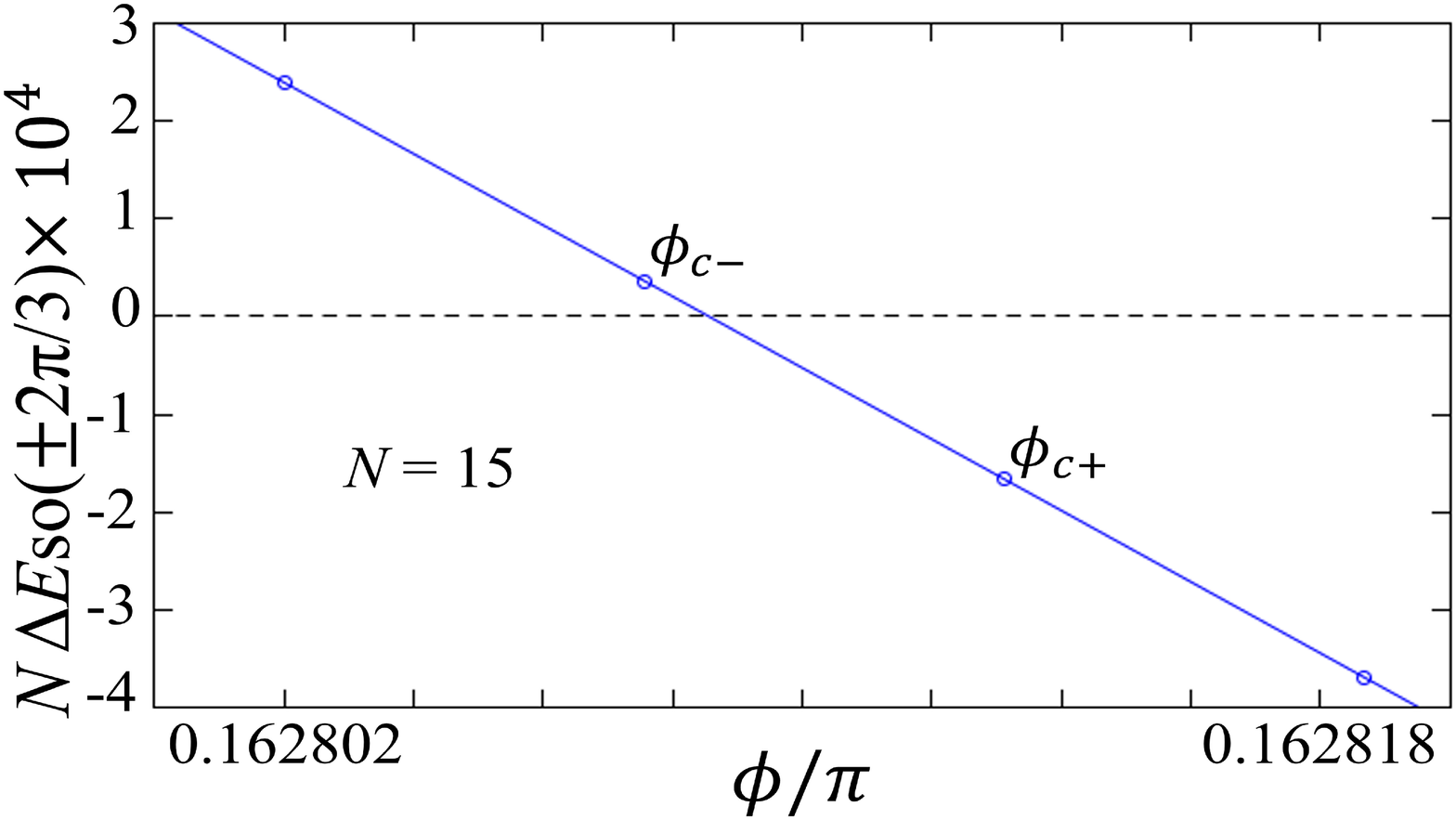}
 \end{center}
 \begin{center}
  \includegraphics[keepaspectratio,scale=0.2]{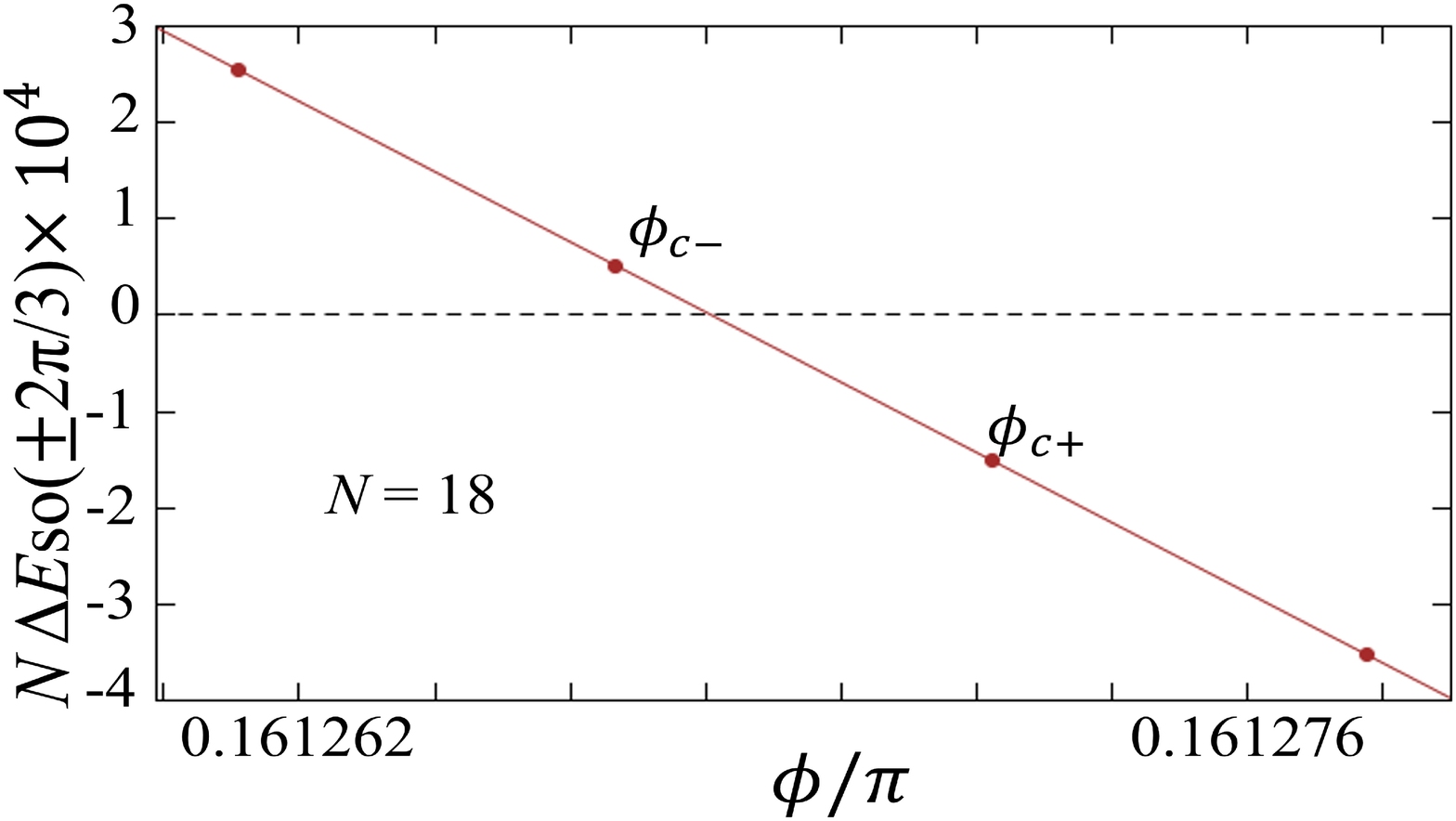}
 \end{center}
 \begin{center}
  \includegraphics[keepaspectratio,scale=0.2]{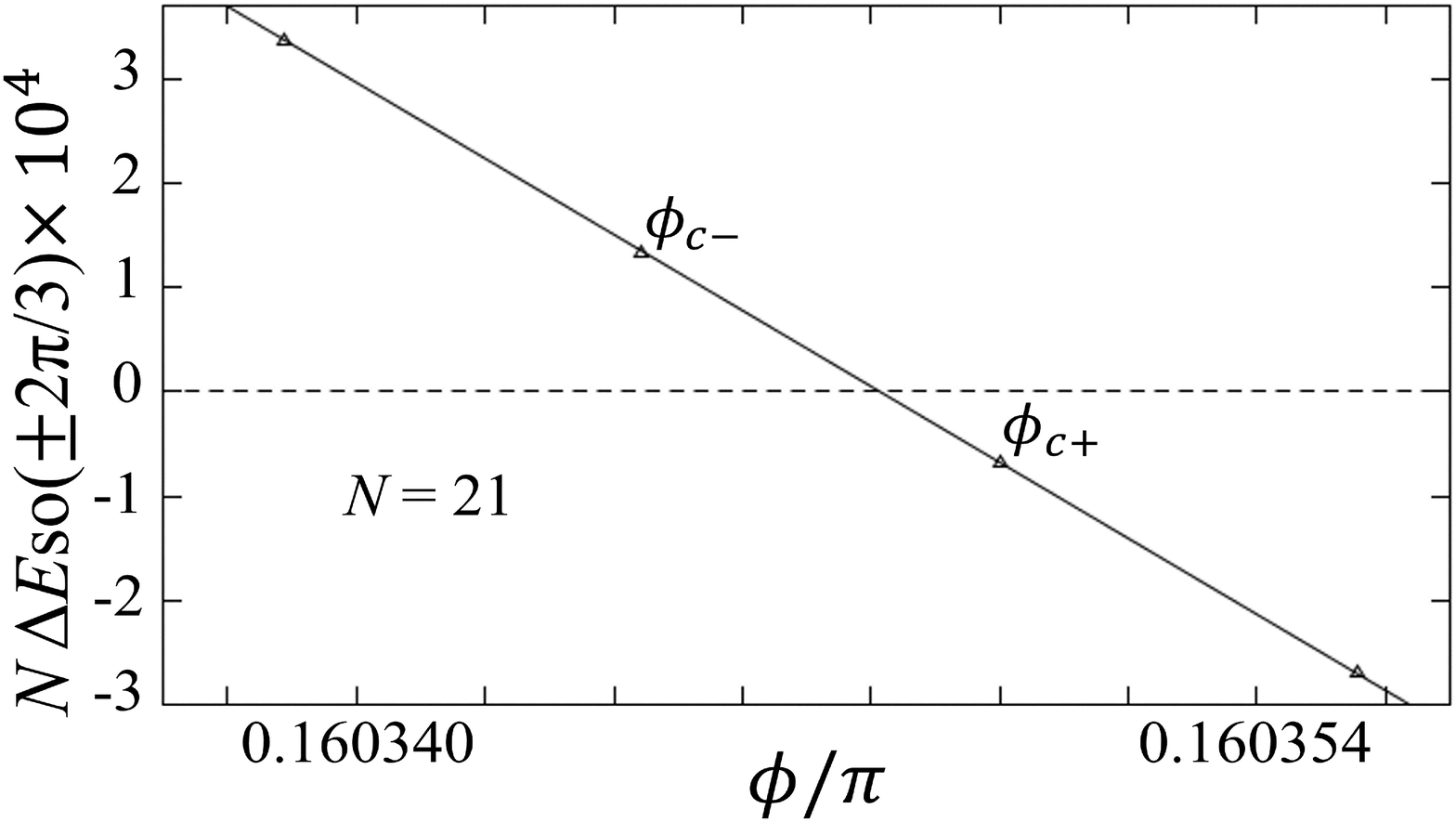}
 \end{center}
        \caption{$N\Delta E_{\mathrm{so}}(\pm 2\pi/3)$ in the vicinity of the phase boundary $\phi_{c}$ with $N=15$-- $21$.}
        \label{fig:esevic}
\end{figure}
In this subsection, we discuss the region in the vicinity of $\phi_{c}$.
Figure \ref{fig:esevic} shows a close-up of Fig. \ref{fig:ese} around the phase boundary $\phi_{c}$. 
Here, we let $\phi_{c+}$ and $\phi_{c-}$ denote parameters $\phi$ in the vicinity of $\phi_{c}$ in the region $\phi>\phi_{c}$ and $\phi<\phi_{c}$ respectively.

\subsubsection{phase boundary}
\begin{figure}[t]
 \begin{center}
  \includegraphics[keepaspectratio,scale=0.23]{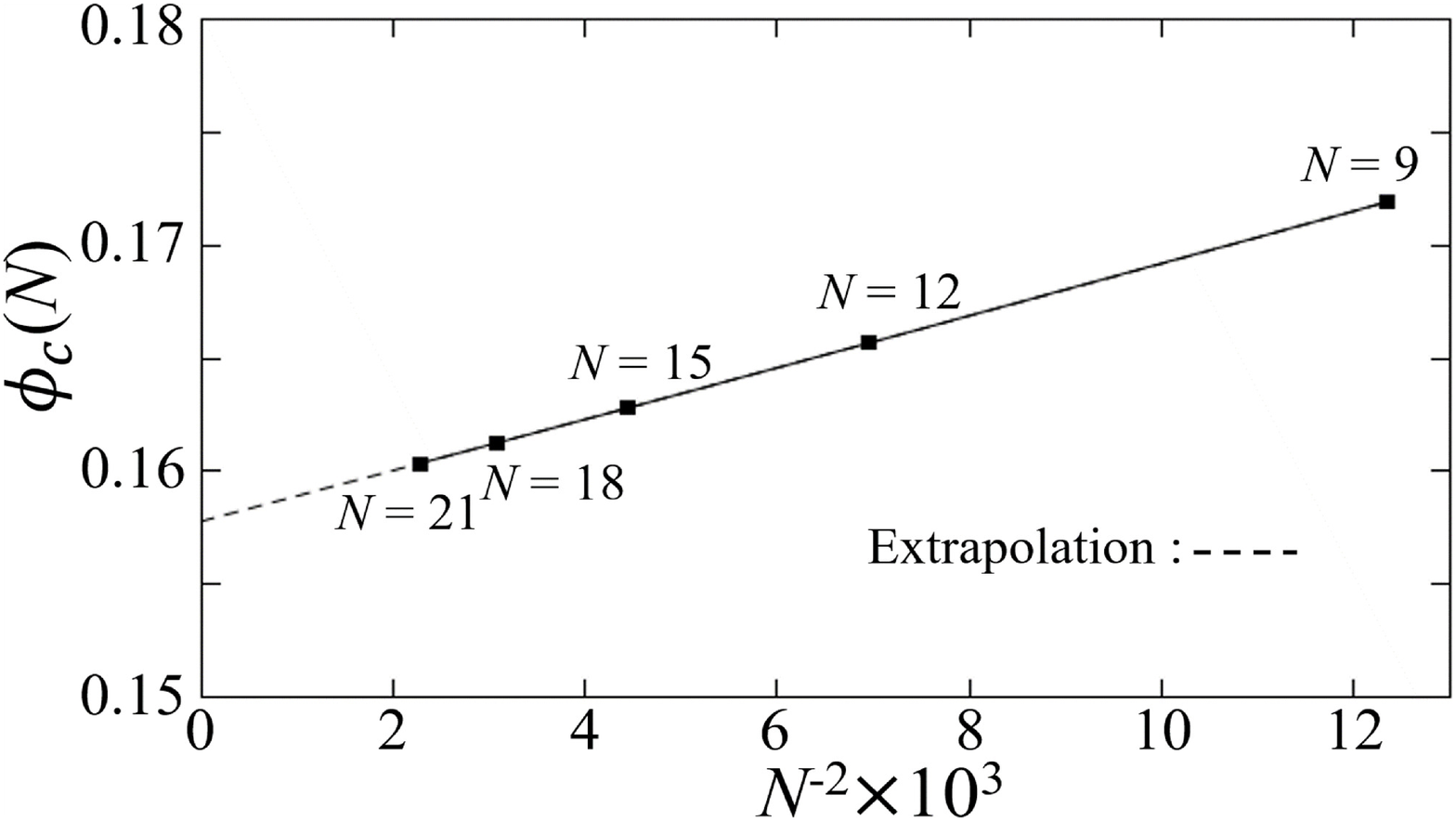}
 \end{center}
	\caption{Phase boundary $\phi_{c}(N)$ as a function of $N^{-2}$ with $N=9$--$21$.}
        \label{fig:phic}
\end{figure}

From the results in Fig \ref{fig:esevic}, we calculate the phase boundary $\phi_{c}$ for each system size utilizing the equation
\begin{eqnarray}
	\frac{N \Delta E_{\mathrm{so}}(\pm 2\pi/3)|_{\phi=\phi_{c+}} }{\phi_{c+} - \phi_{c}} = \frac{N \Delta E_{\mathrm{so}}(\pm 2\pi/3)|_{\phi=\phi_{c-}} }{\phi_{c-} - \phi_{c}}, \notag \\ \label{phicgive}
\end{eqnarray}
where Eq. \eqref{phicgive} is reliable because $\Delta E_{\mathrm{so}}(\pm 2\pi/3)$ depends linearly on $\phi-\phi_{c}$ in the vicinity of $\phi_{c}$.
We plot $\phi_{c}(N)$ in Fig. \ref{fig:phic}.
The correction terms $O \left( N^{-2} \right)$ in Fig. \ref{fig:phic} can be explained by descendant fields of the identity operator with the scaling dimension $x=4$\cite{cardy,cardy0.5,rein,kitazawa}.
Therefore, $\phi_{c}(N)$ behaves as
\begin{eqnarray}
	\phi_{c}(N) = \phi_{c} + A_{1} N^{-2} +A_{2} N^{-4} + O\left(N^{-6} \right), \label{phicn}
\end{eqnarray}
where $A_{1}$ and $A_{2}$ are non-universal constants. 
Fitting the data with a function $\phi_{c}(N) = \phi_{c} + A_{1} N^{-2} +A_{2} N^{-4}$, we obtain 
\begin{eqnarray}
	\frac{\phi_{c}}{\pi} = 0.157794 \pm 0.000001, \label{phicex}
\end{eqnarray}
when we extrapolate $\phi_{c}(N)$ with $N=9$--$21$. 
Therefore, we put $\phi_{c}=0.15779\pi$ later.

\subsubsection{renormalization group and the slope of $N\Delta E_{\mathrm{so}}(\pm 2\pi/3)$ in the vicinity of $\phi=\phi_{c}$}
\label{subsubsec:renvic}
In Fig. \ref{fig:ese}, in the vicinity of $\phi_{c}$, $N\Delta E_{\mathrm{so}}(\pm 2\pi/3)$ seems almost independent of $N$.
We discuss this fact on the basis of the theory of Itoi and Kato\cite{itoi}.

For $|g_{1}(0)| \ll 1$, Eq. \eqref{sol} can be approximated as
\begin{eqnarray}
	g_{1}(l) \approx g_{1}(0). \label{sol3}
\end{eqnarray}
Therefore, substituting Eq. \eqref{sol3} for Eq. \eqref{eso3}, we obtain
\begin{eqnarray}
	N \Delta E_{\mathrm{so}} \left( \pm \frac{2\pi}{3} \right) \approx - \frac{3\pi}{\sqrt{2}} v_{0}(N) g_{1}(0), \label{eso5}
\end{eqnarray}
From Eq. \eqref{eso5}, the behavior of Fig. \ref{fig:ese} in the vicinity of $\phi_{c}$ are explained.
Also, we confirm the expectation Eq. \eqref{g1ex}.

\begin{figure}[t]
 \begin{center}
  \includegraphics[keepaspectratio,scale=0.23]{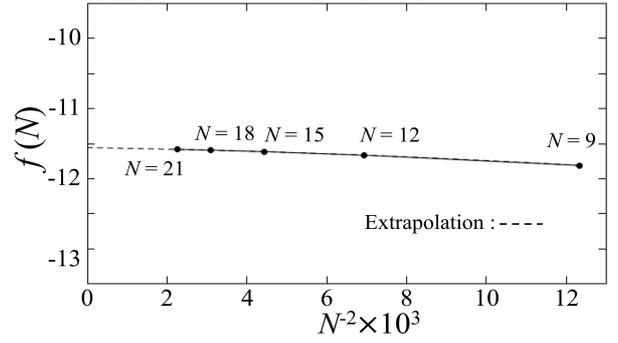}
 \end{center}
        \caption{$f(N)$ as a function of $N^{-2}$ with $N=9$--$21$.}
        \label{fig:slope}
\end{figure}

Next, we derive the constant $C$ in Eq. \eqref{g1ex}.  
To begin with, utilizing Eqs. \eqref{g1ex} and \eqref{eso5}, we obtain a relation
\begin{eqnarray}
	&& N \left[\Delta E_{\mathrm{so}} \left( \left. \pm \frac{2\pi}{3} \right) \right|_{\phi=\phi_{c+}} - \Delta E_{\mathrm{so}} \left( \left. \pm \frac{2\pi}{3} \right) \right|_{\phi=\phi_{c-}} \right] \notag \\
	&& \approx -\frac{3\pi C}{\sqrt{2}} \left[ (\phi_{c+} - \phi_{c}) v_{0}(N)|_{\phi=\phi_{c+}} \right. \notag \\
	&& \hspace{5em} \left. - (\phi_{c-} - \phi_{c}) v_{0}(N)|_{\phi=\phi_{c-}} \right] \notag \\
	&& \approx -\frac{3\pi C}{\sqrt{2}} v_{\mathrm{av}}(N) (\phi_{c+} - \phi_{c-}), \label{eso4} \\
	&& v_{\mathrm{av}}(N) \equiv \frac{v_{0}(N)|_{\phi=\phi_{c+}} + v_{0}(N)|_{\phi=\phi_{c-}}}{2}. \label{vav}
\end{eqnarray}
We plot the slope of $N\Delta E_{\mathrm{so}}(\pm 2\pi/3)$ defined as
\begin{eqnarray}
	&& f(N) \equiv \notag \\
	&& \hspace{1.5em} \frac{N \left[ \Delta E_{\mathrm{so}}(\pm 2\pi/3)|_{\phi=\phi_{c+}} - \Delta E_{\mathrm{so}}(\pm 2\pi/3)|_{\phi=\phi_{c-}} \right]}{\phi_{c+} - \phi_{c-}}, \notag \\ \label{fn}
\end{eqnarray}
in Fig. \ref{fig:slope} as a function of $N^{-2}$.
From Eqs. \eqref{eso4} and \eqref{fn}, the constant $C$ can be written
\begin{eqnarray}
	C \approx -\frac{\sqrt{2}}{3\pi} \cdot \frac{f(N)}{v_{\mathrm{av}}(N)} = -\frac{\sqrt{2}}{3\pi} \cdot \frac{f(\infty)}{v_{\mathrm{av}}(\infty)}. \label{cfn}
\end{eqnarray}
We then calculate $f(\infty)$ and $v_{\mathrm{av}}(\infty)$ with Figs. \ref{fig:v0} and \ref{fig:slope}, and the functions\cite{cardy,cardy0.5,rein,kitazawa}
\begin{eqnarray}
	\hspace{-2em} f(N) &=& f(\infty) + B_{1} N^{-2} + B_{2} N^{-4} + O(N^{-6}), \label{fncorre} \\
	\hspace{-2em} v_{\mathrm{av}}(N) &=& v_{\mathrm{av}}(\infty) + B^{\prime}_{1} N^{-2} + B^{\prime}_{2} N^{-4} + O(N^{-6}), \label{vavcorre}
\end{eqnarray}
where $B_{1}$, $B_{2}$, $B^{\prime}_{1}$, and $B^{\prime}_{2}$ are non-universal constants.
Fitting the data with functions $f(N) = f(\infty) + B_{1} N^{-2} +B_{2} N^{-4}$ and $v_{\mathrm{av}}(N) = v_{\mathrm{av}}(\infty) + B^{\prime}_{1} N^{-2} +B^{\prime}_{2} N^{-4}$, we obtain
\begin{eqnarray}
	f(\infty) &=& -11.5566 \pm 0.0001, \label{finfex} \\
	v_{\mathrm{av}}(\infty) &=& 2.89986 \pm 0.00002, \label{vavinfex}
\end{eqnarray}
when we extrapolate them with $N=9$--$21$.
By substituting Eqs. \eqref{finfex} and \eqref{vavinfex} for Eq. \eqref{cfn}, we obtain $C \approx 0.59797$.

\subsection{Central charge}
\label{subsec:cent}
Here, we calculate the central charge $c$, which characterizes the quantum anomaly and specifies the universality class.
In the critical state of one-dimensional quantum systems under PBCs, the ground-state energy density at $N$ follows\cite{blote,affleck} the equation 
\begin{eqnarray}
	\frac{E_{g}(N)}{N} = \epsilon_{\infty} - \frac{\pi v_{0} c}{6N^{2}}, \label{eg}
\end{eqnarray}
where $\epsilon_{\infty}$ is the ground-state energy density in the case of $N \rightarrow \infty$. 
Note that the central charge has a logarithmic finite-size correction\cite{itoi,cardy1} with the form of $\displaystyle{\frac{4}{9}} (\ln N)^{-3}$.
From Eq. \eqref{eso2}, we estimate $(\ln N)^{-1} \approx 0.32$ in the cases of $N=21$ and $\phi \approx 0$, and thus $\displaystyle{\frac{4}{9}}(\ln N)^{-3} \approx 0.014$, which is small enough compared to $c=2$.
Therefore, we neglect the logarithmic correction in the central charge.
In Eq. \eqref{eg}, one can numerically calculate $E_{g}$ and $v_{0}$, but there remain two constants, $\epsilon_{\infty}$ and $c$, as unknown values. 
By removing the constant term $\epsilon_{\infty}$ in Eq. \eqref{eg}, we thus calculate the effective central charge $c(N)$ as
\begin{eqnarray}
	&&\frac{E_{g}(N)}{N} - \frac{E_{g}(N-3)}{N-3} \notag \\
	&& \hspace{2em} = -\frac{\pi}{6} \left[\frac{v_{0}(N)}{N^{2}} - \frac{v_{0}(N-3)}{(N-3)^{2}} \right] c(N). \label{eg2}
\end{eqnarray}

\begin{figure}[b]
 \begin{center}
  \includegraphics[keepaspectratio,scale=0.23]{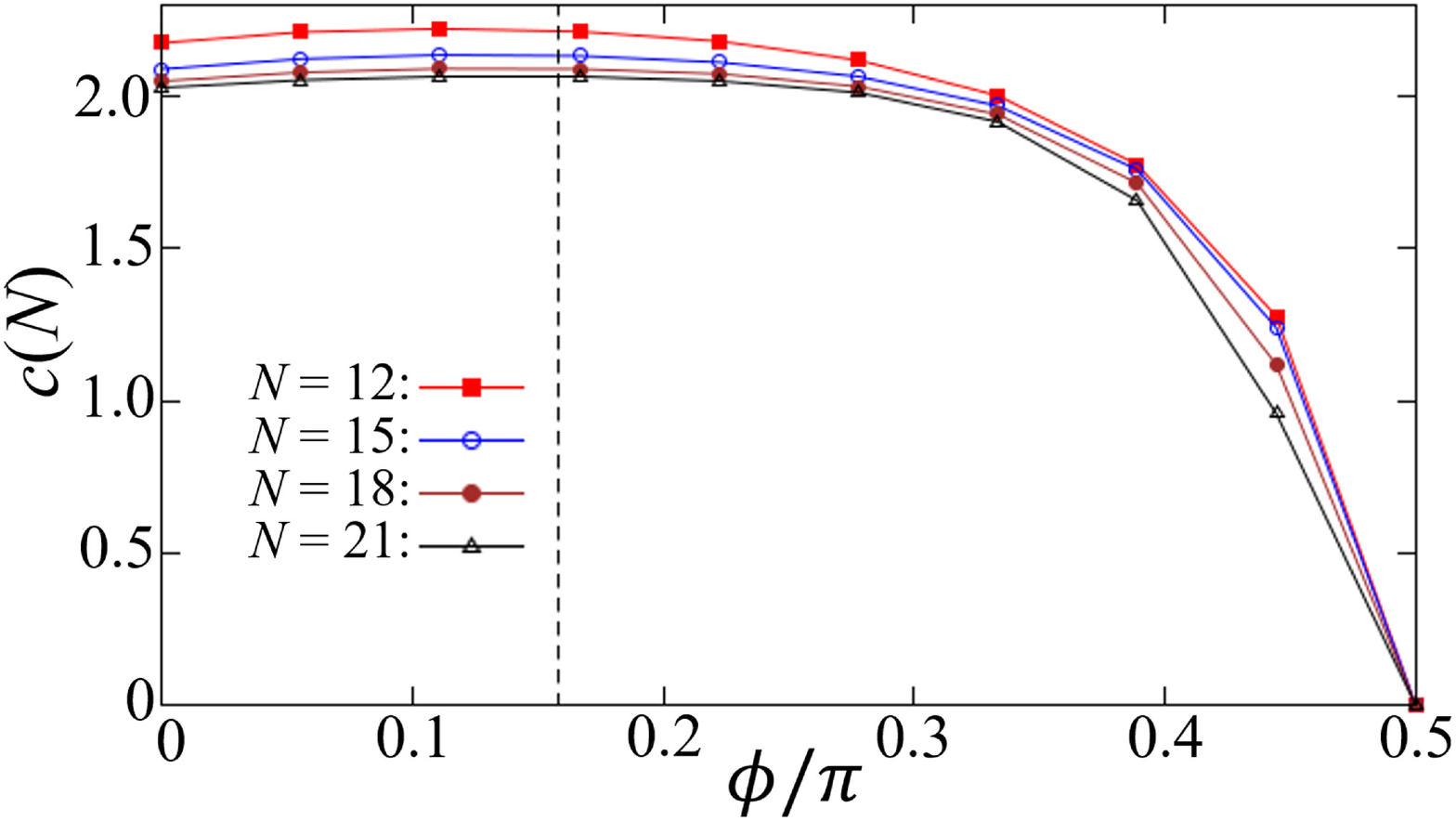}
 \end{center}
	\caption{Effective central charge $c(N)$ as a function of $\phi$ with $N=12$--$21$. The dashed line is $\phi=\phi_{c}$, where we put $\phi_{c}=0.15779\pi$.}
        \label{fig:cph}
\end{figure}

\begin{figure}[t]
 \begin{center}
  \includegraphics[keepaspectratio,scale=0.23]{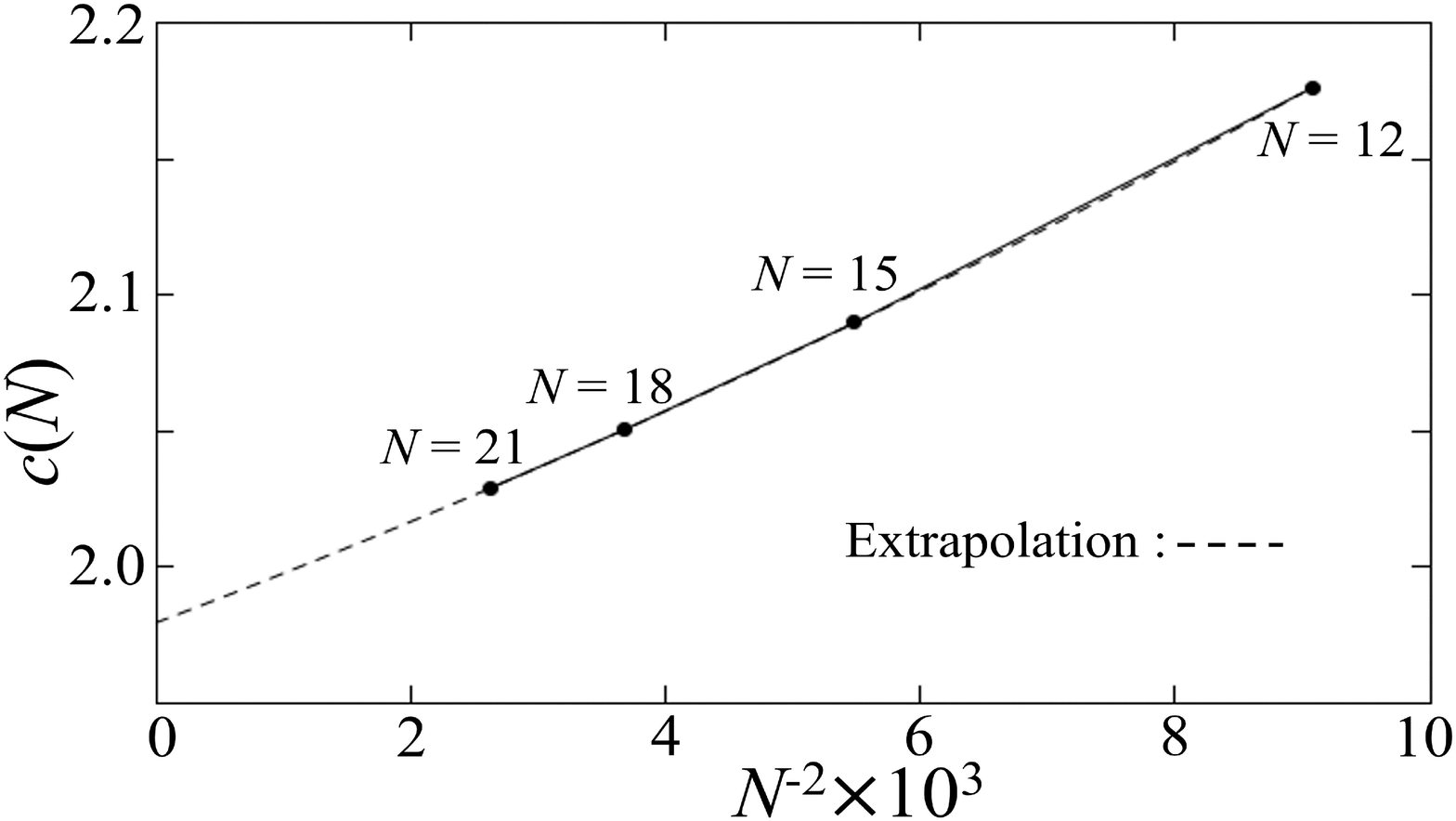}
 \end{center}
	\caption{Effective central charge $c(N)$ with $N=12$--$21$ at $\phi=0$.}
        \label{fig:c-n}
\end{figure}

Figure \ref{fig:cph} shows the effective central charge as a function of $\phi$ for different system sizes, $N=12$--$21$. 
The effective central charge was firstly numerically calculated in the case of the CFT with $c=1$\cite{okamoto}. 
In this article, we investigate the effective central charge utilizing Eq. \eqref{eg2}. 
Equation \eqref{eg} is true only in the case of the massless phase certainly, but we can apply Eq. \eqref{eg2} even to systems in a massive phase as well.
We find that the effective central charges smoothly converge to $c=2$ as $N\rightarrow \infty$ in the region $\phi < \phi_{c}$ (Fig. \ref{fig:c-n}).
In Fig. \ref{fig:c-n}, similarly to Eq. \eqref{phicn}, we extrapolate the effective central charge $c(N)$ as\cite{cardy,cardy0.5,rein,kitazawa}
\begin{eqnarray}
	c(N) &=& c + C_{1} (N-3/2)^{-2} +C_{2} (N-3/2)^{-4} \notag \\
		&& \hspace{7em} + O\left((N-3/2)^{-6}\right), \label{cf}
\end{eqnarray}
where $C_1$ and $C_2$ are constants.
Then, we obtain $c = 1.97963 \pm 0.00006$ when we extrapolate the $c(N)$ with the function Eq. \eqref{cf} with $N=12$--$21$.

On the other hand, in the region $\phi > \phi_{c}$, the effective central charge shows a decline as $\phi$ approaches $\pi/2$. 
Moreover, around $\phi=\phi_{c}$, there are no sharp decline of $c(N)$ because of the logarithmic correction\cite{cardy1} $O\left(\left(\ln N \right)^{-3} \right)$.

Comparing our numerical results in Figs. \ref{fig:cph} and \ref{fig:c-n} with Zamolodchikov's $c$-theorem\cite{zamolodchikov} and the theory of Itoi and Kato\cite{itoi}, we find that the region $\phi < \phi_{c}$ is illustrated by the CFT with $c=2$ (critical phase), whereas the region $\phi > \phi_{c}$ is a massive phase. 

\subsection{Scaling dimension}
\label{subsec:x}
\begin{figure}[b]
 \begin{center}
  \includegraphics[keepaspectratio,scale=0.23]{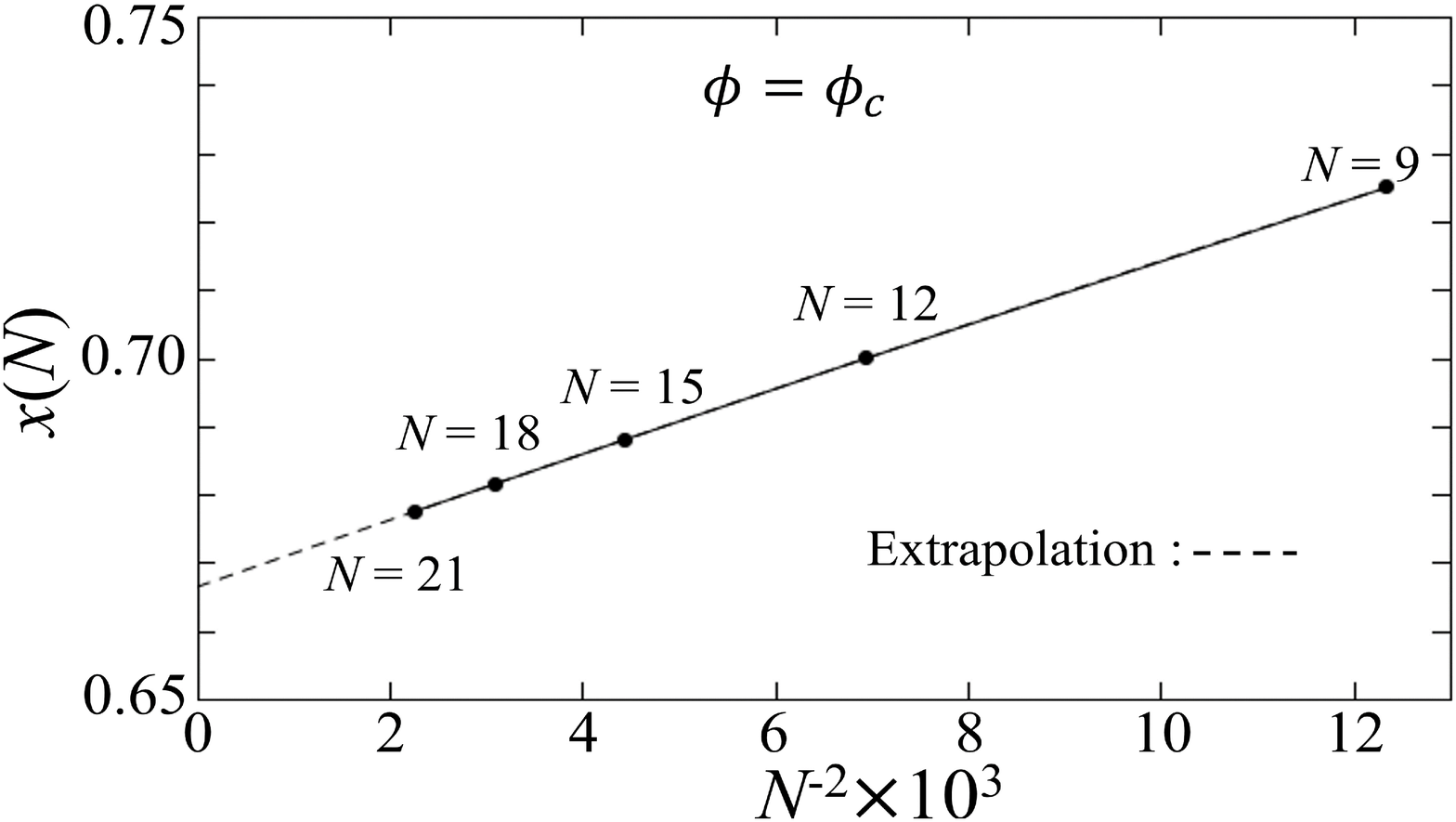}
 \end{center}
	\caption{Effective scaling dimension $x(N)$ with $N=9$--$21$ at $\phi=\phi_{c}$.}
        \label{fig:x-n}
\end{figure}

In this subsection, we calculate the scaling dimension $x$, which is one of the critical exponents, at the boundary ($\phi=\phi_{c}$). 
Here, we rewrite elementary excitation energy at a certain $S_T$, Eq. \eqref{estp}, as
\begin{eqnarray}
	\Delta E_{S_T} \left( \pm \frac{2\pi}{3} \right)=\frac{2\pi v_{0}}{N}\left[x+d_{S_T}g_{1}(l) \right],\label{de}
\end{eqnarray}
where we put $0 < g_{1}(l) \ll 1$.
We derive the scaling dimension $x$ by removing $d_{S_T}g_{1}(l)$ in Eq. \eqref{de} from the values in Eq. \eqref{cst} as
\begin{eqnarray}
	&& \frac{1}{9} \left[\Delta E_{0} \left( \pm \frac{2\pi}{3} \right) + 3\Delta E_{1} \left( \pm \frac{2\pi}{3} \right) + 5 \Delta E_{2} \left( \pm \frac{2\pi}{3} \right) \right] \notag \\
	&& \hspace{14em} = \frac{2\pi v_{0}(N)}{N}x(N). \label{x}
\end{eqnarray}

Similarly to Eq. \eqref{phicn}, the effective scaling dimension $x(N)$ behaves as\cite{cardy,cardy0.5,rein,kitazawa}
\begin{eqnarray}
        x(N) = x + D_{1} N^{-2} +D_{2} N^{-4} + O\left( N^{-6} \right), \label{xf}
\end{eqnarray}
where $D_{1}$ and $D_{2}$ are constants.

Figure \ref{fig:x-n} shows the effective scaling dimension $x(N)$ at $\phi=\phi_{c}$. 
We obtain $x = 0.666616 \pm 0.000004$ when we extrapolate the $x(N)$ with the function Eq. \eqref{xf} in the cases of $N = 9$--$21$.

These numerical results at $\phi=\phi_{c}$ point are in line with the scaling dimension $x=2/3$ of the SU$(3)_{1}$ WZW model\cite{wess,witten,witten2}.

\section{CONCLUSION AND DISCUSSION}
\label{sec:con}
We have investigated the model Eq. \eqref{muls} to clarify the critical behavior when changing the parameter $\phi$. 
First of all, we find that soft modes appear at the wave number $q=0$, $\pm 2\pi/3$ in both cases of $\phi=0$ and $\phi=\pi/2$. 
Secondly, we find that the phase transition is caused by a marginal operator.
In other words, there occurs a phase transition between the massless (marginally irrelevant) phase and  the $\mathbb{Z}_{3}$ symmetry broken (marginally relevant) phase at $\phi_{c}=0.15779\pi$. 
In the case of SU($\nu$) symmetric systems, where the  critical phenomena are caused by the marginal operator, there are only a few numerical researches\cite{okamoto,chepiga1,chepiga2} combining renormalization group discussion.
Thirdly, by investigating the central charge and the scaling dimension, we find that the region $\phi < \phi_{c}$ is a critical phase with $c=2$ and $x=2/3$ whereas the region $\phi < \phi_{c}$ is a massive phase.
From these numerical results, we conclude that there occurs a phase transition at $\phi=\phi_{c}$ between the TL phase $(\phi < \phi_{c})$ and the trimer phase $(\phi > \phi_{c})$.

Here, we discuss several correlation functions in the SU(3) symmetric TL phase to discuss the critical behavior.
First of all, we discuss the spin correlation function and the spin-quadrupolar correlation function, which correspond to the energy of the triplet state $\Delta E_{1}(\pm 2\pi/3)$ and the quintuplet state $\Delta E_{2}(\pm 2\pi/3)$ respectively.
It is expected\cite{itoi,stou,lauch} to be
\begin{eqnarray}
	&&\left \langle \hat{\bm{S}}_{i} \cdot \hat{\bm{S}}_{i+r}  \right \rangle = \left \langle \hat{Q}^{\mu\nu}_{(i)} \hat{Q}_{(i+r)\mu\nu}  \right \rangle \notag \\ 
	&& \hspace{2em}\propto \cos \left(\frac{2\pi}{3}r \right) r^{-4/3} \left(\ln r \right)^{2/9}, \label{sqcorre} \\
	&&\hat{Q}^{\mu\nu}_{(i)} \equiv \frac{1}{2} \left\{\hat{S}^{\mu}_{i}, \hat{S}^{\nu}_{i} \right\} - \frac{2}{3}\delta^{\mu\nu}, \label{sqorder}
\end{eqnarray}
from $c=2$ and $x=2/3$. 
Here, $\hat{Q}^{\mu\nu}_{(i)}$ is the spin-quadrupolar order parameter at site $i$, which is symmetric and traceless. 
Secondly, Schmitt et al. proposed\cite{schm} an order parameter corresponding to the singlet state $\Delta E_{0}(\pm 2\pi/3)$, which is defined as
\begin{eqnarray}
	\hat{T}_{i} \equiv \hat{T}^{\mathrm{P}}_{i} - \left \langle \hat{T}^{\mathrm{P}}_{i} \right \rangle, \hspace{1em} \hat{T}^{\mathrm{P}}_{i} \equiv \left| \{i,j,k \} \right \rangle \left \langle \{i,j,k \} \right|, \label{ordertri}
\end{eqnarray}
where the state vector of $\left| \{i,j,k \} \right \rangle$ is the same as $\{ \circ \circ \circ \}$, Eq. \eqref{trimer2}, and we put $j\equiv i+1$ and $k\equiv i+2$.
The correlation function of $\hat{T}_{i}$ is expected\cite{itoi} to be
\begin{eqnarray}
	\left \langle \hat{T}_{i} \hat{T}_{i+r} \right \rangle \propto \cos \left(\frac{2\pi}{3}r \right) r^{-4/3} \left(\ln r \right)^{-16/9}. \label{tprocorre} 
\end{eqnarray}

In the $\mathbb{Z}_{3}$ ordered phase, the correlation functions behave as follows.
As for the spin correlation function and the spin-quadrupolar correlation function, it is expected to be
\begin{eqnarray}
	\hspace{-2em} \left \langle \hat{\bm{S}}_{i} \cdot \hat{\bm{S}}_{i+r}  \right \rangle = \left \langle \hat{Q}^{\mu\nu}_{(i)} \hat{Q}_{(i+r)\mu\nu}  \right \rangle \propto \cos \left(\frac{2\pi}{3}r \right) e^{-r/\xi}, \label{sqcorrez3}
\end{eqnarray}
where the correlation length $\xi$ is defined in Eq. \eqref{xi}.
Next, the correlation function of Eq. \eqref{ordertri} behaves\cite{schm} as
\begin{eqnarray}
	&& \left \langle \hat{T}_{i} \hat{T}_{i+r} \right \rangle \propto \cos \left(\frac{2\pi}{3}r \right). \label{tprocorrez3}
\end{eqnarray}
which characterizes a trimer long-range order.
Also, in the vicinity of the phase boundary $\phi_{c}$, i.e., $\xi \rightarrow \infty$, these three functions are almost the same.

Recently, controllable quantum many-body systems have been realised in ultracold atoms in an optical lattice. 
We also believe that our numerical results can be applied to experiments and quantum simulations in such systems. 
Especially, quantum many-body systems have SU($\nu$) ($\nu$: integer) symmetry in ultracold strontium ($^{87}$Sr) atoms\cite{desalvo} and ytterbium ($^{173}$Yb) atoms\cite{taie} in an optical lattice. 
These atomic systems are expected to be applied to quantum information processing. 
Our results will be a part of a basic theory in realising quantum information processing.

\section*{ACKNOWLEDGEMENTS}
We are grateful to S. Moriya for constructing a useful algorithm, which we modify for our numerical calculations.

\appendix
\section{SU(3) AND ITS REPRESENTATION}
\label{sec:greiter}
In this section, we review SU(3) and the representation of SU(3) on the basis of the paper\cite{greiter} and the book\cite{group}, corresponding to the trimer state.
\subsection{Gell-Mann matrices}
\label{subsec:gell}
The Gell-Mann matrices, the basis of SU(3) group, are defined as
\begin{eqnarray*}
\lambda^1=\left(
    \begin{array}{ccc}
      0 & 1 & 0 \\
      1 & 0 & 0 \\
      0 & 0 & 0
    \end{array}
  \right),
\lambda^2=\left(
    \begin{array}{ccc}
      0 & -i & 0 \\
      i & 0 & 0 \\
      0 & 0 & 0
    \end{array}
  \right),
\lambda^3=\left(
    \begin{array}{ccc}
      1 & 0 & 0 \\
      0 & -1 & 0 \\
      0 & 0 & 0
    \end{array}
  \right), \vspace{0.5em}\\
\lambda^4=\left(
    \begin{array}{ccc}
      0 & 0 & 1 \\
      0 & 0 & 0 \\
      1 & 0 & 0
    \end{array}
  \right),
\lambda^5=\left(
    \begin{array}{ccc}
      0 & 0 & -i \\
      0 & 0 & 0 \\
      i & 0 & 0
    \end{array}
          \right),
\lambda^6=\left(
    \begin{array}{ccc}
      0 & 0 & 0 \\
      0 & 0 & 1 \\
      0 & 1 & 0
    \end{array}
  \right), \vspace{0.5em}\\
\lambda^7=\left(
    \begin{array}{ccc}
      0 & 0 & 0 \\
      0 & 0 & -i \\
      0 & i & 0
    \end{array}
  \right),
\lambda^8=\frac{1}{\sqrt{3}}\left(
    \begin{array}{ccc}
      1 & 0 & 0 \\
      0 & 1 & 0 \\
      0 & 0 & -2
    \end{array}
  \right).
\end{eqnarray*}
They satisfy the commutation relations $[\lambda^{a},\lambda^{b}] = 2 f^{abc} \lambda^{c}$ and are normalized as $\mathrm{tr}(\lambda^{a}\lambda^{b}) = 2\delta_{ab}$.
The structure constants $f^{abc}$ are totally antisymmetric and the non-vanishing values are given by $f^{123}=i$, $f^{147}=f^{246}=f^{257}=f^{345}=-f^{156}= -f^{367}= i/2$, $f^{458}=f^{678}=i\sqrt{3}/2$.

\subsection{Representation of SU(3)}
\label{subsec:rep}
\begin{figure}[t]
 \begin{center}
  \includegraphics[keepaspectratio,scale=0.28]{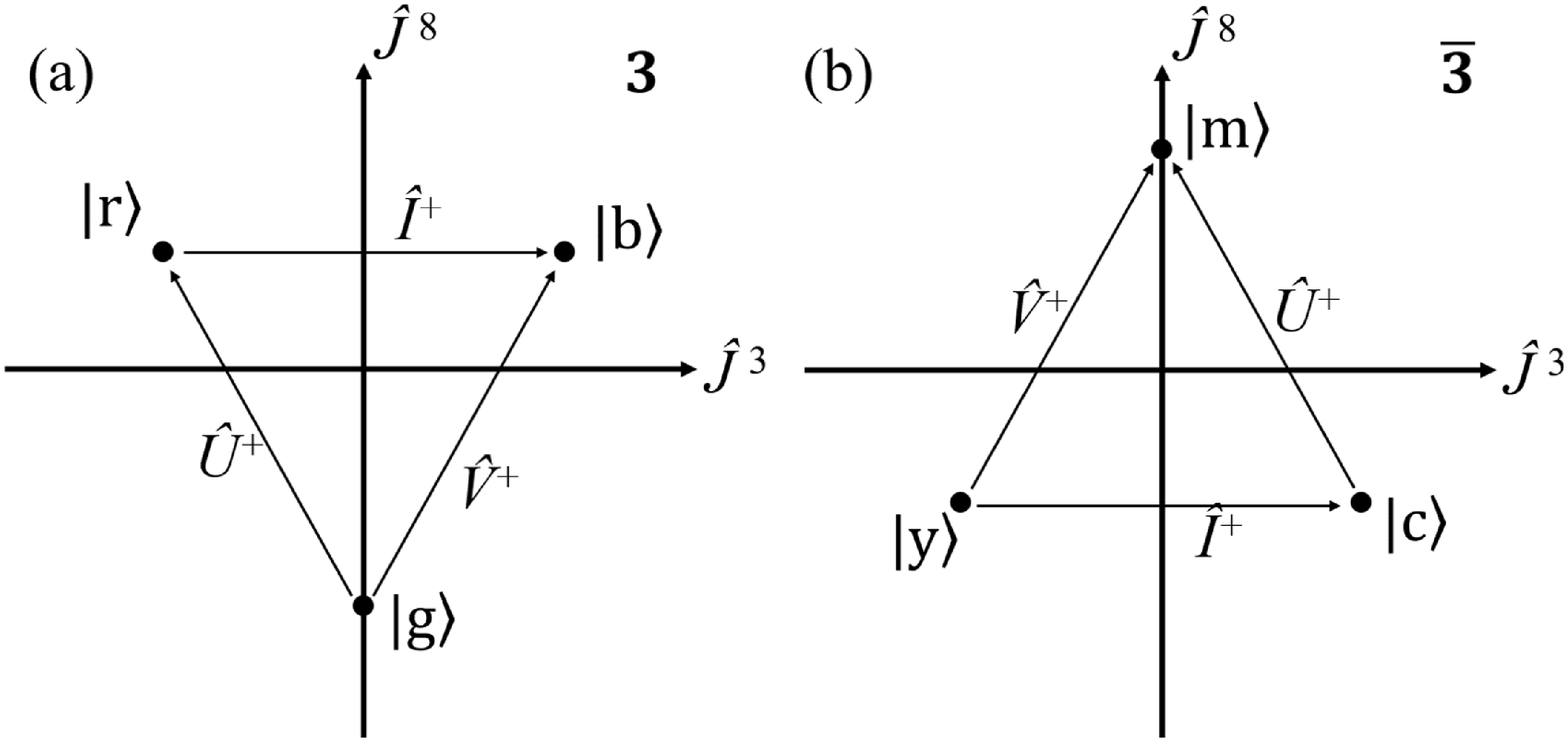}
 \end{center}
        \caption{Weight diagrams of the fundamental representations, (a) $\bm{3}$ and (b) $\bar{\bm{3}}$.}
        \label{fig:3}
\end{figure}
\begin{figure}[b]
 \begin{center}
  \includegraphics[keepaspectratio,scale=0.23]{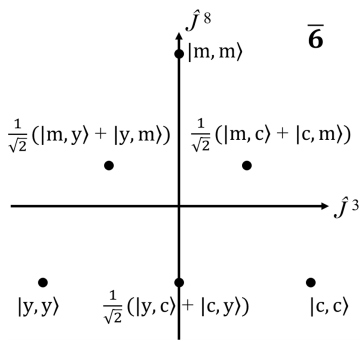}
 \end{center}
        \caption{Weight diagram of the representation $\bar{\bm{6}}$.}
        \label{fig:6}
\end{figure}
Generators $\hat{J}^{a}$, Eq. \eqref{gen}, satisfy the algebra
\begin{eqnarray}
	\left[ \hat{J}^{a},\hat{J}^{b} \right] = f^{abc}\hat{J}^{c}. \label{alge}
\end{eqnarray}
For the group SU(3), there are two diagonal generators $\hat{J}^{3}$ and $\hat{J}^{8}$.
Also, the other generators define the ladder operators as
\begin{eqnarray}
	\hat{I}^{\pm} \equiv \hat{J}^{1} \pm i \hat{J}^{2}, \notag \\
	\hat{U}^{\pm} \equiv \hat{J}^{6} \pm i \hat{J}^{7}, \notag \\
	\hat{V}^{\pm} \equiv \hat{J}^{4} \pm i \hat{J}^{5}, \notag
\end{eqnarray}
respectively.
The algebra Eq. \eqref{alge} is realized by the $\hat{J}^{a}$'s, which gives the fundamental representation $\bm{3}$ of SU(3) illustrated in Fig. \ref{fig:3}.
The weight diagram in Fig. \ref{fig:3} instructs us about the eigenvalues of the diagonal generators as well as the actions of the ladder operators of the basis states. Here, we put $|\mathrm{b} \rangle \equiv |1 \rangle$, $|\mathrm{r} \rangle \equiv |0 \rangle$, $|\mathrm{g} \rangle \equiv |-1 \rangle$.

According to Greiter and Rachel\cite{greiter}, a spin-1 $\circ$, Eq. \eqref{triplet} $\{\circ\circ \}$, and the trimer $\{\circ\circ\circ \}$ are expressed by the representations $\bm{3}$, $\bar{\bm{3}}$, and $\bm{1}$ respectively.
The fundamental representation $\bar{\bm{3}}$ is also illustrated in Fig. \ref{fig:3}, where we define
\begin{eqnarray}
        && |\mathrm{m} \rangle \equiv \frac{1}{\sqrt{2}} \left(|1,0 \rangle - |0,1 \rangle \right), \\
        && |\mathrm{c} \rangle \equiv \frac{1}{\sqrt{2}} \left(|1,-1 \rangle - |-1,1 \rangle \right), \\
        && |\mathrm{y} \rangle \equiv \frac{1}{\sqrt{2}} \left(|0,-1 \rangle - |-1,0 \rangle \right).
\end{eqnarray}
Also, the trimer states on four sites in Eq. \eqref{auxi} can be expressed as
\begin{eqnarray}
&&	\{\circ\circ\circ \} \circ \hat{=} \bm{1} \otimes \bm{3} = \bm{3}, \\
	&&	\{\circ\circ \}\{\circ\circ \} \hat{=} \bar{\bm{3}} \otimes \bar{\bm{3}} = \bm{3} \oplus \bar{\bm{6}},
\end{eqnarray}
where the weight diagram of $\bar{\bm{6}}$ is depicted in Fig. \ref{fig:6}

\section{RENORMALIZATION-GROUP EQUATION}
\label{sec:itoi}
In this section, we review the RG calculation by Itoi and Kato\cite{itoi}  to investigate the critical behaviour around the transition point $\phi=\phi_{c}$ in this paper.

First of all, we let $x_{0}$ denote the time in the system and $x_{1}$ be the position of the field. 
We then define $z$ and $\bar{z}$ as
\begin{eqnarray}
z \equiv x_{0} + i x_{1}, \,\,\,\,\,\, \bar{z} \equiv x_{0} - i x_{1}.
\end{eqnarray}
We define the action $\hat{\mathcal{A}}$ as
\begin{eqnarray}
        \hat{\mathcal{A}} \equiv \hat{\mathcal{A}}_{\mathrm{SU}(3)_{1}} + g_{1} \int \frac{d^{2}z}{2\pi} \hat{\Phi}^{(1)} \left( z , \bar{z} \right), \label{asu}
\end{eqnarray}
where $\hat{\mathcal{A}}_{\mathrm{SU}(3)_{1}}$ is the action of the free fields in the SU$(3)_{1}$ WZW model\cite{wess,witten,witten2}. 
$\hat{\Phi}^{(1)}$ is an operator of the marginal field with SU(3) symmetry and chiral $\mathbb{Z}_{3}$ symmetry. 
The scaling variable $g_{1}$ is a perturbational parameter. 
The system remains SU$(3)$ symmetric regardless of the value of $g_{1}$. 
According to Itoi and Kato\cite{itoi}, the renormalization-group equation for the scaling variables becomes
\begin{eqnarray}
	\frac{d g_{1}(l)}{dl} = \frac{3}{2\sqrt{2}} g_{1}^{2}(l). \label{rg3}
\end{eqnarray}

\end{document}